\documentclass[aps,prb,twocolumn,superscriptaddress,
showpacs,preprintnumbers,amsmath,amssymb
,floatfix
]{revtex4}

\usepackage{graphicx}
\usepackage{dcolumn}
\usepackage{bm}

\begin{document}

\title{
Time-Reversal Coherent Control in Nanoplasmonics
}
\author{Xiangting Li}
\email{xiangtingli@yahoo.com}
\affiliation{Department of Physics and Astronomy, Georgia State
University, Atlanta, GA 30303, USA
}
\affiliation{Department of Physics, Shanghai Jiaotong University,
Shanghai 200240, China
}
\author{Mark I. Stockman}
\email{mstockman@gsu.edu}
\homepage{http://www.phy-astr.gsu.edu/stockman}
\affiliation{Department of Physics and Astronomy, Georgia State
University, Atlanta, GA 30303, USA
}

\date{\today}

\begin{abstract}
We introduce an approach to determining the required waveforms to
coherently control the optical energy localization in plasmonic
nanosystems. This approach is based on the impulsive localized
excitation of the nanosystem and time reversal of the generated far-zone
field at a single point with one polarization. Despite strong
interaction and significant dephasing and dissipation in metal plasmonic
systems, and incompleteness of this time reversal, the proposed approach
proves to be very efficient in controlling the nanoscale optical fields.
Possible applications include nanoscale spectroscopy and
photomodification, ultradense memory, and information processing on the
nanoscale.
\end{abstract}


\maketitle

One of the challenging problems is control of the nanoscale localization
of optical excitation energy in nanosystems. The principal physical
origin of this problem is that the wavelength of the optical radiation is
orders of magnitude larger than the size of a nanoscale system.
Therefore the optical radiation cannot be focused into a nanoscale spot,
i.e., it does not posses spatial degrees of freedom on the nanoscale.
However, it does possess the temporal degrees of freedom, which is the
same, in a different language, as the frequency or phase degrees of
freedom. The idea to use the phase modulation of the optical radiation
to control the nanoscale localization of the optical energy has been
proposed \cite{Stockman:2002_PRL_control} and led to a significant
subsequent development both theoretical 
\cite{Phys_Rev_B_69_054202_2004_Stockman_Bergman_Kobayashi_Coherent_Control,
Stockman_Hewageegana_Nano_Lett_2005_V_Shape_Coherent_Control,
Brixner_et_al_PRL_2005_Nanoscopic_Ultrafast_Spectroscopy,
Sukharev_Seideman_JCP_2006_Coherent_Control_on_Nanoscale} and
experimental
\cite{Kubo_Onda_Petek_Sun_Jung_Kim_Nano_Lett_2005_Two_Pulse_Coherent_Control,
Aeschlimann_Bauer_Bayer_Brixner_et_al_Nature_446_301_2007_Nanooptical_Adaptive_Control}.
In symmetric systems, polarization of the excitation has been an
efficient degree of freedom complementing the phase and amplitude
modulation. 
\cite{Brixner_et_al_PRL_2005_Nanoscopic_Ultrafast_Spectroscopy,
Sukharev_Seideman_JCP_2006_Coherent_Control_on_Nanoscale,
Aeschlimann_Bauer_Bayer_Brixner_et_al_Nature_446_301_2007_Nanooptical_Adaptive_Control}
Coherent control of quantum and nanoscopic systems is a
powerful tool of defining pathways of optically-excited processes in
them.
\cite{Kurizki_Shapiro_Brumer_PRB_1989_Coherent_Control_of_Currents_in_Semiconductors,
Brumer_Shapiro_Ann_Rev_Phys_Chem_1992_Laser_Control_of_Molecules,
Rabitz_et_al_Science_288_824_2000_Quantum_Control,
Geremia_Rabitz_PRL_89_263902_2002_Optimal_Hamiltonian_Identification,
Nguyen_Dey_Shapiro_Brumer_JPCA_108_7878_2004,
Shapiro_Brumer_Phys_Rep_415_195_2006_Quantum_Control_of_Dynamics}

One of the fundamental problems in the coherent control is the solution
of its ``inverse problem'': finding an optical waveform that sends the
controlled system along the required excitation pathway. One approach to
this problem is the adaptive optimum control that has proved successful in a
wide class of problems.
\cite{Rabitz_et_al_Science_288_824_2000_Quantum_Control,
Geremia_Rabitz_PRL_89_263902_2002_Optimal_Hamiltonian_Identification,
Sukharev_Seideman_JCP_2006_Coherent_Control_on_Nanoscale,
Aeschlimann_Bauer_Bayer_Brixner_et_al_Nature_446_301_2007_Nanooptical_Adaptive_Control} 
However, in the adaptive algorithms, it is sometimes
difficult to interpret the obtained complicated waveforms. A
problem in theoretical investigations is that implementations of the
adaptive algorithms are often computationally costly.

In this Letter we propose and theoretically investigate a novel approach
to finding an efficient optical pulse controlling a nanosystem. It
is based on an idea of time reversal (or, back propagation). We
start with an initial state of the plasmonic nanosystem where a
localized excitation is prepared at a desired nanosite. Then we solve
the direct problem of evolution and propagate the fields to the far
zone. At some moment $t_r$, we time-reverse the far-zone pulse and send
it back to the system. If the system were completely time reversible,
then its evolution would back-track itself causing the concentration of
energy at the desired site in time $t_r$ after the instance of the time
reversal. This idea is significantly based on the previous acoustic and
microwave studies.
\cite{Derode_Tourin_de_Rosny_Tanter_Yon_Fink_PRL_2003_Ultrasonic_Time_Reversal,
Fink_et_al_PRL_92_193904_2004_Time_Reversal_of_EM_Waves,
Lerosey_de_Rosny_Tourin_Fink_Science_315_1120_2007_Microwave_Time_Reversal_Subwavelength_Focusing}

However, there are principal differences of surface plasmon (SP)
eigenmodes in metal nanosystems from the reverberating, leaky modes in
acoustics and microwaves of Refs.\
\onlinecite{Derode_Tourin_de_Rosny_Tanter_Yon_Fink_PRL_2003_Ultrasonic_Time_Reversal, 
Fink_et_al_PRL_92_193904_2004_Time_Reversal_of_EM_Waves,
Lerosey_de_Rosny_Tourin_Fink_Science_315_1120_2007_Microwave_Time_Reversal_Subwavelength_Focusing}, 
which make controlling the SPs much more
difficult. First, due to the strong interaction of dipolar excitations
on the nanoscale, the SPs in nanosystems form chaotic eigenmodes that can
be delocalized over the entire nanosystem.
\cite{Stockman_PRL_1997_Eigenmode_Chaos,
Stockman_PRE_1997_Eigenode_Chaos_Correlations} This phenomenon manifests
itself as correlated ``hot spots'' of local fields. Obviously, both the
delocalization of the SPs and their chaoticity (including high
sensitivity to parameters) hamper the ability to control them. Second
important difference is that the real metals in the plasmonic spectral
region are lossy. \cite{Johnson:1972_Silver} 
Therefore, SPs have finite life times and
are not exactly time-reversible. The third problem, which is common to
the acoustics, microwaves, and plasmonics, is that the far field
does not contain the full information of the internal state of the
systems: the evanescent fields are vanishingly small and lost in the far
zone. Moreover, the far-zone field is normally measured in a limited
number of points with incomplete polarization
information (in the extreme case, at one point with a single
polarization). Therefore, the full time-reversal of the 
field is impossible. These serious problems notwithstanding,
as we show below in this Letter, the time reversal of the far-zone field even
at a single point with a single polarization produces a signal that is
capable of providing an excellent control of the optical field
nano-localization.
 
Turning to the theory, we consider
a nanostructured system consisting of metal and dielectric with the
permittivities $\varepsilon_m$ and $\varepsilon_d$, respectively. The
entire size of this system is assumed to be much less than the
wavelength of the excitation radiation. Therefore we can use the
quasistatic spectral theory. \cite{Stockman:2001_PRL_Localization,
Phys_Rev_B_69_054202_2004_Stockman_Bergman_Kobayashi_Coherent_Control}
We start with an optical dipole $\mathbf d(\mathbf r_0, t)$ localized at
a point $\mathbf r_0$ at the metal surface whose density is $\mathbf
P(\mathbf r)=\delta(\mathbf r- \mathbf r_0) \mathbf d(\mathbf r_0, t)$,
and the dependence on time $t$ is a short pulse containing frequencies
$\omega$ centered around the carrier frequency $\omega_0$.

This initial oscillating dipole causes the appearance of local fields 
$\mathbf E^L(\mathbf r, t)$ in the system that are given by 
\cite{Kneipp_Moskovits_Kneipp_SERS_Springer_Verlag_2005}
\begin{equation}
E^L_\alpha({\bf r},\omega)=\frac{4\pi}{\varepsilon_d}\,
G^r_{\alpha\beta}({\mathbf r}, {\mathbf r}_0;\omega)\,
d_{\beta}({\mathbf r_0},\omega)~,
\label{E_r_0}
\end{equation}
where the Greek subscripts denote vector indices with summation over
repeated indices implied.
Here and below, by indicating a frequency variable $\omega$ we imply the
Fourier transform of the corresponding temporal function; e.g., $\mathbf
E(\mathbf r,\omega)=\int_{-\infty}^\infty \mathbf E(\mathbf
r,t)\exp(i\omega t)\mathrm d t$. Retarded dyadic Green's function 
$G^r_{\alpha\beta}$ is expressed in terms of the corresponding scalar
Green's function $\bar G^r$:
\begin{equation}
G^r_{\alpha\beta}({\bf r}, {\bf r}^\prime;\omega)=
\frac{\partial^2}{\partial r_\alpha\partial r^\prime_\beta}
\bar G^r({\bf r}, {\bf r}^\prime;\omega)~.
\label{diadic_Greens}
\end{equation}
This is given as an expansion over the
eigenfunctions $\varphi_n$ and eigenvalues $s_n$ of the SP
eigenproblem \cite{Stockman:2001_PRL_Localization,
Phys_Rev_B_69_054202_2004_Stockman_Bergman_Kobayashi_Coherent_Control,
Kneipp_Moskovits_Kneipp_SERS_Springer_Verlag_2005}
\begin{equation}
\bar G^r({\bf r}, {\bf r}^\prime; \omega)=
\sum_n \frac{\varphi_n({\bf r})\,\varphi_n({\bf r}^\prime)^\ast}
{s(\omega)-s_n}~,~~~ s(\omega)=\frac{\varepsilon_d}{\varepsilon_d-
\varepsilon_m}~.
\label{Greens_expansion}
\end{equation}

\begin{figure}
\centering
\includegraphics[width=.45\textwidth]
{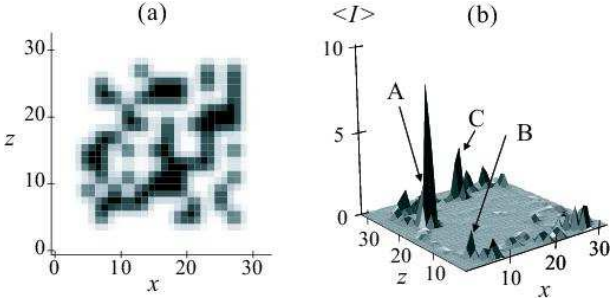}
\caption{\label{Sample.eps}
(a) Geometry of the random planar composite used in the computations,
shown in
the $xz$ projection. The unit length for the axes is 1 nm, but can be
scaled. The system is scalable in the limits allowed by
the quasistatic approximation. (b) Average (over pulse time)
intensity of local fields is displayed distributed over the surface of the
nanosystem. The fields magnitude is shown in the units of the
excitation pulse amplitude, which is set as 1, and whose length is 5 fs
with mean frequency $\omega_0=1.2$ eV.
}
\end{figure}

The total radiating dipole moment of the nanosystem $\mathbf D$, which
defines the field in the far zone, is the seed dipole $\mathbf
d$ plus the dipole of the entire system induced by field $\mathbf E^L$
(\ref{E_r_0}) that renormalizes and enhances it (the antenna effect). 
It is given in the frequency domain as
\cite{Kneipp_Moskovits_Kneipp_SERS_Springer_Verlag_2005}
\begin{equation}
D_\alpha(\omega)=\left[\delta_{\alpha\beta}%
-\frac{1}{s(\omega)}g_{\beta\alpha}({\bf r}_0;\omega)\right]
d_{\beta}({\bf r_0},\omega)~, ~~~
\label{D_d}
\end{equation}
\begin{equation}
g_{\alpha\beta}({\bf r},\omega)=
\int_V G^r_{\alpha\beta}({\bf r}, {\bf r}^\prime;\omega)
\Theta({\bf r}^\prime)\, {\rm d}^3 r^\prime~,
\label{g}
\end{equation}
where $\Theta({\bf r}^\prime)$ is the characteristic function equal to 1 
when ${\bf r}^\prime$ belongs to the metal and to 0 otherwise. To
complete the solution of the inverse problem of the coherent control, we
find the time-reversed radiating dipole moment 
\begin{equation}
\mathbf D^T(t)=\int_{-
\infty}^\infty D^\ast(\omega)\exp(-i\omega t)\mathrm d\omega/(2\pi)~.
\label{D_T}
\end{equation}
The field generated by this oscillating dipole in the far zone is then
used to excite the nanosystem. 

\begin{figure}
\centering
\includegraphics[width=.45\textwidth]
{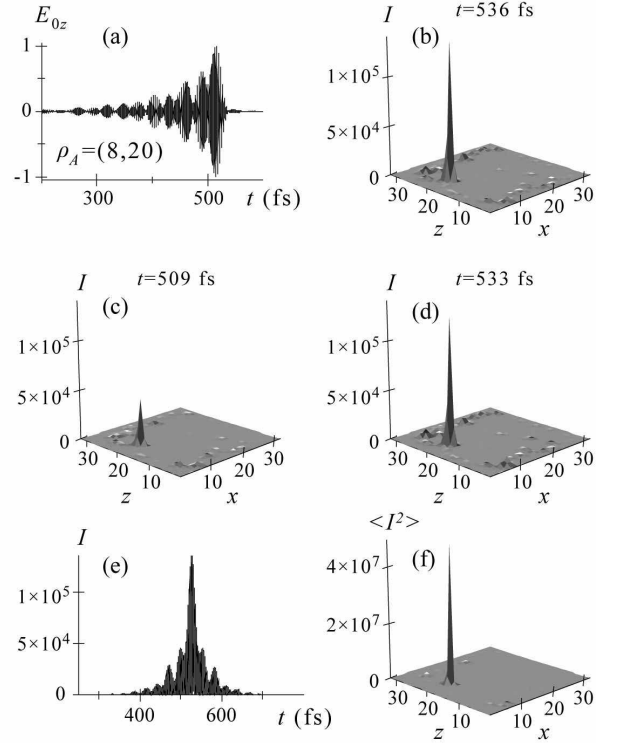}
\caption{\label{PositionA_zdirection.eps}
(a) Excitation field at the system as computed by time-reversal
for the initial dipole at the A point
(normalized to 1 at the maximum). (b)-(d) Distributions of the local
field intensity at the points A, B, and C, correspondingly, calculated
for the instances of their respective maxima. (e) Time evolution of the
local intensity at the targeted point A. (f) Distribution of the time-averaged 
squared intensity over the surface of the nanosystem. Units of field and
intensity are arbitrary but consistent for all the panels.
}
\end{figure}

The final step is the solution of the direct problem, i.e. finding the 
the local field $\mathbf E(\mathbf r,t)$ in the system that is excited
by the time-reversed {\it uniform} field $\mathbf E^T(t)$. 
This can be expressed in the Fourier domain as
\begin{equation}
E_\alpha({\bf r},\omega)=\left[\delta_{\alpha\beta}+%
g_{\alpha\beta}({\bf r},\omega)\right]
E^T_{\beta}(\omega)~.
\label{E_g}
\end{equation}
If the time-reversal is efficient in solving the inverse problem, 
then the local field, Fourier-transformed to the time domain, 
should demonstrate the concentration at the initial 
site $\mathbf r_0$ in time $t_r$ after the reversal.

As a numerical illustration, we consider a random planar composite which
is a 4-nm thick layer of silver \cite{Johnson:1972_Silver} in vacuum. 
This layer is 50 percent randomly filled with $2\times2\times2
~\mathrm{nm^3}$ unit cells, as shown in Fig.\
\ref{Sample.eps} (a). The total size of this composites is
$32\times4\times32 ~\mathrm{nm^3}$. In the quasistatic approximation, the
system is scalable, as long as its total size is still much less the
light wavelength.

There are {\it a priori} limitations on the coherent control in
the plasmonic nanosystems. In particular, the local field energy can
only be localized at the sites where the eigenmodes with frequencies
within the bandwidth of the excitation pulse are localized. To get an
idea where in the nanosystem such a localization takes place, we apply
a very short, 5-fs duration, unmodulated Gaussian excitation pulse whose
carrier frequency $\omega_0=1.2$ eV is in the window of the least dephasing of the
SPs. \cite{Bergman_Stockman:2003_PRL_spaser} The resulting local field
intensity averaged over the pulse time, is shown in Fig.\ \ref{Sample.eps} (b).
Among the many hot spots of the local fields seen in this panel, we
choose three peaks marked as A, B, and C, whose $\rho=(x,z)$ coordinates
at the surface are $\rho_A=(8,20)$, $\rho_B=(5,5)$, and
$\rho_C=(20,28)$, correspondingly. In all cases the time dependence of
the initial dipole $\mathbf d(\mathbf r_0,t)$ has been set as a pulse
with a 20 fs Gaussian envelope and $\omega_0=1.2$
eV. The initial dipole was polarized in the $z$-direction. Separate
computations for this dipole $x$-polarized have given very similar
results (not shown), which is due to the strong depolarization effect of
the present random nanostructure. This property of random, complex
nanosystems is in sharp contrast to the polarization-driven control for
a symmetric system. 
\cite{Aeschlimann_Bauer_Bayer_Brixner_et_al_Nature_446_301_2007_Nanooptical_Adaptive_Control} 
The time-reversed excitation field in
the far zone has been calculated from the dipole (\ref{D_T}). Its
maximum amplitude on the system has been normalized to 1 to make the
comparison of the plasmonic enhancements easier. 

\begin{figure}
\centering
\includegraphics[width=.45\textwidth]
{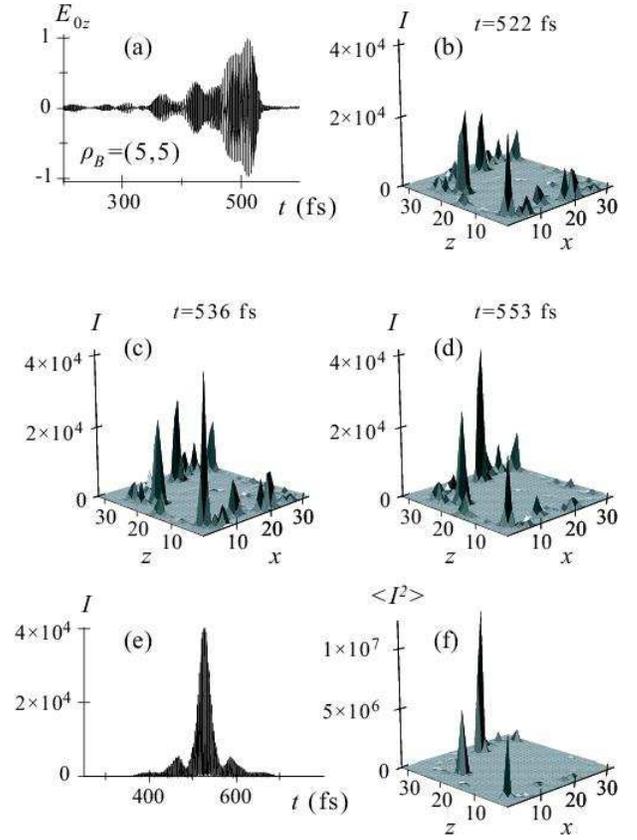}
\caption{\label{PositionB_zdirection.eps}
The same as in Fig.\ \ref{PositionA_zdirection.eps} but for the initial
dipole at the B point.
}
\end{figure}

When the initial dipole is at the point A, the
calculated  time-reversed electric field incident on the system 
is shown in Fig.\ \ref{PositionA_zdirection.eps} (a) where $t_r=536$ fs.
We can see that
this field is dramatically different from the seed 20-fs 
Gaussian-envelope polarization that generated it (after the time reversal).
The pulse is
relatively long, with bursts of fields and their revivals, accompanied
by the general decay due to the dephasing. This pulse looks very similar
to the pulses obtained by the time reversal in acoustics and microwaves
\cite{Derode_Tourin_de_Rosny_Tanter_Yon_Fink_PRL_2003_Ultrasonic_Time_Reversal,
Fink_et_al_PRL_92_193904_2004_Time_Reversal_of_EM_Waves,
Lerosey_de_Rosny_Tourin_Fink_Science_315_1120_2007_Microwave_Time_Reversal_Subwavelength_Focusing}
where this behavior is due the reverberations of the propagating waves
repeatedly reflected from the boundaries and inhomogeneities of the
system; the decay of the signal is due to the leakage of the wave energy
from those open systems. However, the similarity stops here, because in
our case the observed beatings are due to the interference of the
localized, non-propagating (quasistatic) SPs eigenmodes; the decay is
due to their dephasing occurring both due to the multitude and randomness
of the eigenmode frequencies involved, and also due to the dephasing of
the metal electron polarization as described by $\mathrm{Im}\,\varepsilon_m$. 

In Figs.\ \ref{PositionA_zdirection.eps} (b)-(d) we display the
local field intensity $I(\mathbf r,t)= \left|\mathbf E(\mathbf
r,t)\right|^2$ distributed over the surface of the nanosystem for three
moments of time where the intensities at sites A, B, and C are maximum,
correspondingly. Note that in all these cases the intial dipole is at
the A point. The maximum concentration of energy at the A site is almost
perfect [panel (b)]; it is reached at $t=536$ fs, which coincides with
the expected time $t_r$ (the end of the excitation pulse). Comparing
to the case of an ummodulated pulse [Fig.\ \ref{Sample.eps} (b)], the
excitation of the other peaks is almost completely suppressed. Even when
the ``undesired'' peaks B and C go through their temporal maxima [panels
(c) and (d)], the targeted peak A is still dominant. Not only the
spatial structure of the local fields is highly concentrated. Also the
temporal evolution of the local field intensity at the A point shown in
Fig.\ \ref{PositionA_zdirection.eps} (e) is restored almost completely
to its initial Gaussian envelope (though with some pedestal). Finally,
we display in Fig.\ \ref{PositionA_zdirection.eps} (f) the
time-integrated square of the intensity, which describes the
distribution of the two-photon electron-emission current as measured, e.
g., by a photoemission electron microscope (PEEM). 
\cite{Kubo_Onda_Petek_Sun_Jung_Kim_Nano_Lett_2005_Two_Pulse_Coherent_Control,
Aeschlimann_Bauer_Bayer_Brixner_et_al_Nature_446_301_2007_Nanooptical_Adaptive_Control}
Such a current is almost ideally concentrated at the targeted point A.

\begin{figure}
\centering
\includegraphics[width=.45\textwidth]
{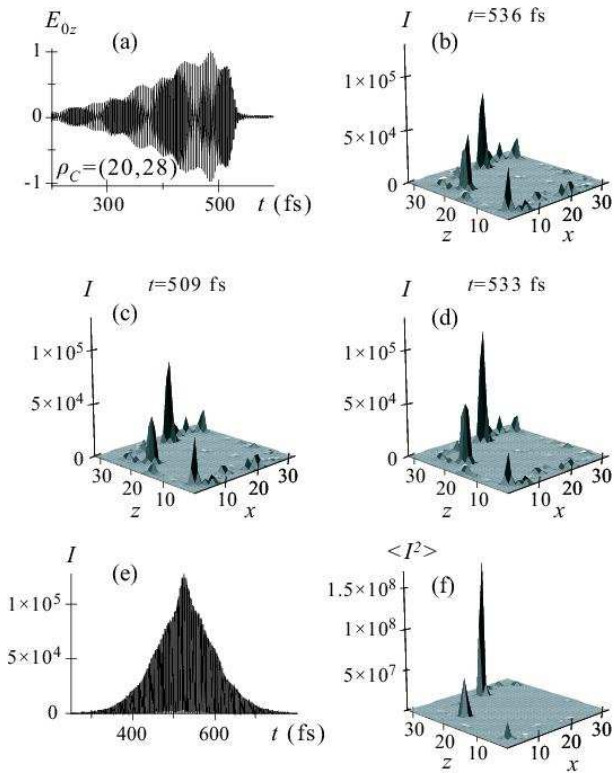}
\caption{\label{PositionC_zdirection.eps}
The same as in Fig.\ \ref{PositionA_zdirection.eps} but for the initial
dipole at the C point.
}
\end{figure}

Similar case for the initial dipole at the B point is illustrated in
Fig.\ \ref{PositionB_zdirection.eps}. The excitation, time-reversed
pulse [panel (a)] is significantly different from the previous case: the
beatings are obviously much less frequent. The spatial distributions
of the local field intensity displayed in panels (b)-(d) reveal that
the initially weak peak B [cf.\ Fig.\ \ref{Sample.eps} (b)] is
relatively very much enhanced. It reaches its maximum [panel (c)] at
$t=536$ fs coinciding with the expected (back-tracking) time $t_r$, 
where it is the largest peak]. This is certainly a success of the coherent
control. Moreover, the time evolution of the local fields at the B site
[panel (e)] shows an excellent temporal concentration and reproduction
of the seed 20 fs Gaussian pulse. However, as panel (d) shows, the maximum
magnitude of the peak at the C site is comparable, though smaller, than
that of the maximum B peak. It is also important that the initially
strongest peak A [Fig.\ \ref{Sample.eps} (b)] is significantly
suppressed. The time-averaged nonlinear current [panel (f)]
is nevertheless dominated by the C
site, which is due to the very long-lived local fields at that site.
Thus the temporal concentration at the targeted point B
driven by the time-reversed field is sharp and transient in time.

For the C point as targeted, illustrated in Fig.\
\ref{PositionC_zdirection.eps}, the excitation, time-reversed pulse
[panel (a)] is very long, lacking strong beatings, which implies a weak
dephasing. The C peak certainly dominates the temporal dynamics reaching
its maximum panel (d)] at $t=533$ fs, which is just one period of
oscillations shifted from the back-tracking time $t_r=556$ fs. The C
peak also dominates the nonlinear current [panel (f)]. The temporal
dynamics at the targeted C point [panel (e)] shows a rather broadened
peak, but its very center exhibits some narrow spike at $t\approx t_r$.
Overall, the time-reversal coherent control is very efficient at the
selective concentration of the excitation energy at this point.

To briefly conclude, we have introduced an efficient approach to solving
an and important and formidable problem of the coherent control
of the local optical energy distribution in plasmonic nanosystems. We
have start with a localized dipole producing a short pulse of optical
oscillations at a targeted site of the nanosystem. The field of this
dipole with one polarization at a single point in the far zone is
time-reversed and used as an excitation pulse. We have shown above that
despite the significant problems in time reversing a lossy, strongly
interacting plasmonic nanosystems using incomplete information, it is
still possible to efficiently concentrate the energy of ultrafast
optical fields at the targeted nano-site. Thus, the time reversal
provides a powerful method to solve this fundamental problem of the
coherent control. This method can be used either as an alternative to or
in combination with the adaptive coherent control where it provides the
initial pulse. This proposed approach can be used for controlling the
ultrafast local optical dynamics in nanosystems for a variety of
applications, including superdense and ultrafast optical memory and
computing on the nanoscale, ultrafast local spectroscopy and
photochemistry on the nanoscale, and others.

This work was supported by grants from the Chemical Sciences, Biosciences and
Geosciences Division of the Office of Basic Energy Sciences, Office of
Science, U.S. Department of Energy, a grant CHE-0507147 from NSF, 
and a grant from the US-Israel BSF. 



\end{document}